\input harvmac.tex
\vskip 0.5in
\Title{\vbox{\baselineskip12pt
\hbox to \hsize{\hfill}
\hbox to \hsize{\hfill OUTP-00-57P}
\hbox to \hsize{\hfill KEK-TH-735}
\hbox to \hsize{\hfill hep-th/0012128}}}
{\vbox{\centerline{World Sheet Logarithmic CFT in AdS Strings,} 
\vskip 0.3in
\centerline {Ghost-Matter Mixing and M-theory.}
\vskip 0.3in
{\vbox{\centerline{}}}}}
\centerline{ Ian I.  Kogan\footnote{$^\dagger$}
{i.kogan1@physics.oxford.ac.uk}}
\centerline{\it Theoretical Physics,  Oxford University, 1 Keble Road}
\centerline{\it Oxford, OX13NP, UK}
\medskip
\centerline{ Dimitri Polyakov\footnote{$^\dagger$}
{polyakov@post.kek.jp}}
\centerline{\it High Energy Accelerator Research Organization (KEK)}
\centerline{\it Tsukuba, Ibaraki 305-0801, Japan }
\vskip .2in
\centerline {\bf Abstract}
We discuss several closely related concepts in the NSR formulation of 
 superstring theory.  We demonstrated that recently proposed NSR model 
 for superstrings on  $AdS_5 \times S^5$  is described by the
world-sheet logarithmic conformal field theory (LCFT). The origin of
LCFT on a world-sheet is closely connected to the matter-ghost mixing
in the structure of a brane-like vortex operators. We suggest a
dynamical origin of M theory as a  string theory with an  extra
dimension  given  by  bosonised superconfonformal ghosts.
\vskip .1in

{\bf Keywords:} 
{\bf PACS:}$04.50.+h$;$11.25.Mj$. 
\Date{December 2000}
\vfill\eject
\lref\wil{K.G.Wilson, Phys.Rev.D10:2445-2459,1974}
\lref\kts{R.Kallosh, A.Tseytlin,{\bf J.High Energy Phys.9810:016,1998}}
\lref\pe{I.Pesando,{\bf JHEP11(1998)002}}
\lref\mets{R.Metsaev, A.Tseytlin,{\bf Phys.Lett.B436:281-288,1998}}
\lref\ampf{S.Gubser,I.Klebanov, A.M.Polyakov, 
  Phys.Lett.B428,105.}
\lref\ampstringconf{ A. M. Polyakov,  talk at STRINGS'97
   Nucl.Phys.Proc.Suppl. 68 (1998) 1-8 }
\lref\amps{A.M.Polyakov,  Nucl.Phys.B486:23-33,1997. }
\lref\ampt{A.M.Polyakov,hep-th/9809057}
\lref\wit{E.Witten{\bf Adv.Theor.Math.Phys.2:253-291,1998}}
\lref\gibbons{G.Gibbons,M.Green,M.Perry
{\bf Phys.Lett. B370 (1996)}}
\lref\gibb{G.Gibbons, P.Townsend{\bf Phys.Rev.Lett.71(1993),5223}}
\lref\kleb{S.Gubser,A.Hashimoto,I.Klebanov,J.Maldacena,
{\bf Nucl.Phys.B472:231-248,1996}}
\lref\myself{D.Polyakov, Phys.Lett. {\bf B}  469 (1999) 103}
\lref\self{D.Polyakov,{hep-th/0005094}}
\lref\gs{M.Green,J.Schwarz,{\bf Phys.Lett.B136 (1984) 367}}
\lref\sw{J.Schwarz{\bf Nucl.Phys.B226(1983) 269}}
\lref\krr{R.Kallosh,J.Rahmfeld, A.Rajaraman,hep-th/9805217}
\lref\mts{R.Metsaev,A.Tseytlin,hep-th/9806095}
\lref\polchinski{J.Polchinski,{\bf Phys.Rev.Lett.75:4724-4727,1995}}
\lref\ramf{R.Kallosh, J.Rahmfeld, hep-th/9808038}
\lref\KM{I.I. Kogan and N.E. Mavromatos,  Phys. Lett. {\bf B375}
 (1996), 111;  hep-th/9512210.}
\lref\KMW{I. I. Kogan, N. E. Mavromatos and J. F. Wheater,
Phys.Lett. {\bf B387} (1996) 483, hep-th/9606102}
\lref\BK {A. Bilal and I.I. Kogan, PUPT-1482, hep-th/9407151 
(unpublished); Nucl. Phys. {\bf B 449}, (1995) 569; hep-th/9503209}
\lref\Gur { V. Gurarie, Nucl. Phys. {\bf B 410} (1993), 535}
\lref\CKT {J.S. Caux, I.I. Kogan and A. Tsvelik,
Nucl. Phys. {\bf  B 466} (1996), 444; hep-th/9511130}
\lref\susy { I.I. Kogan, talks at Triangle Meeting, Copenhagen,
June 2000, ``30 Years of SUSY'' Minneapolis, October 2000, to be
published}
\lref\disorder {
 J. Cardy, cond-mat/9911024,
 V. Gurarie and A. W. W. Ludwig, cond-mat/9911392,
M. R. Rahimi Tabar, cond-mat/0002309}
\lref\huffel{P.Damgaard, H.Huffel,{\bf Phys. Rep. B52 (1987)227}}
\lref\amp{A.S.Gubser, I.Klebanov, A.M.Polyakov, Phys.Lett.B428:105}
\lref\malda{J.Maldacena, Adv. Theor. Math. Physics 2 (1998) 231}
\lref\wit{E.Witten, Adv. Theor. Math. Phys.2 (1998): 253}
\lref\verlinde{E.Verlinde, H.Verlinde hep-th/9912018}
\lref\dhoker{E. D'Hoker, D.Freedman, A.Matusis, Nucl. Phys. B 456:96
 (1999)}
\lref\amp{A.M. Polyakov, hep-th/0006132}
\lref\sen{A. Sen, hep-lat/0011073, hep-th/9802051}
\lref\mwitten{ E. Witten,  Nucl. Phys. B 443 (1995) 85}
\lref\mtown{ P. K. Townsend, Phys. lett. B 350 (1995) 184}
\lref\fieldstrings{ A.M.Polyakov, 
"Gauge Fields and Strings", Harwood Academic Publishers, (1987).}
\lref\gsw{ M.B.Green, J.H.Schwarz and E.Witten, "Superstring Theory",
 Cambridge University Press, (1987).}
\lref\jp{ J. Polchinski,``String Theory'', 
Cambridge University Press, (1998).} 
\lref\jpdbranes { J. Polchinski, Phys. Rev. Lett. 75 (1995), 4724}
\lref\dimacohomology{ D.Polyakov,Phys.Rev. D57 (1998) 2564-2570} 
\lref\tsvelik{ M.J. Bhaseen et al, cond-mat/0012240}
\lref\bars{ I. Bars, hep-th/9608061, hep-th/0008164}

\centerline{\bf Introduction}  
 The question of what is a proper  non-perturbative formulation of
String Theory is one of the most important. Last decade we had an
enormous progress in our understanding of different aspects of
non-perturbative formulation of String Theory (see for example  very
interesting lectures \refs{\amp}, \refs{\sen} and references therein) - 
but  we still walk along the shore of an ocean. One of the most
remarkable suggestion was that instead of superstring theory in 10
dimensions we have so called M theory in 11 dimensions and instead of
(super)string as a fundamental object we have a supermembrane - $M2$
brane  \refs{\mwitten},\refs{\mtown}. 
These ideas  lead to very elegant unification of all existing
string theories (and 11-dimensional supergravity). However  the complete
firstly quantized theory of  $M2$ brane (and it dual $M5$ brane) 
is still unknown - contrary to (super)strings \refs{\fieldstrings}, 
\refs{\gsw}, \refs{\jp}. Another remarkable development was the
discovery of D-branes in superstring theory \refs{\jpdbranes}. From
the point of view of M theory one can get type IIA D-branes  from 
 $M2$ and $M5$ branes.  The third remarkable progress was the
discovery of  AdS/CFT correspondence \refs{\malda},\refs{\ampf},
\refs{\wit}   and related   picture of  closed strings in
$d$-dimensional gauge theories  (loops of glue)  propagating actually
in a $d+1$- dimensional space \refs{\ampstringconf}. The emergence of
extra dimension is due to Liouville filed, i.e. one has to start with
non-critical $d$-dimensional string and dynamical Liouville field
plays the role of a new  dimension transforming $d$-dimensional space
into $d+1$-dimensional. 

 It will be nice if one can make such a trick and get extra dimension
moving us from 10 to 11, it will be also nice to find inside string
theory objects which can play the role of $M2$ and $M5$ branes and it
will be nice to understand what kind of  world-sheet dynamics  we must 
 have to describe these objects.
 In this paper we are addressing these questions. In the next section
we are discussing the new brane vertex operators in NSR formulation of 
 superstring theory which amusingly enough can actually play the role
of creation operators of $M2$ and $M5$ branes. We show that world
sheet conformal field theory describing them is rather unusual and
finally we demonstrate that there is an {\bf extra}  bosonic field
 in superstring theory -  superconformal ghost which  indeed produces
an extra eleventh dimension!

\centerline{\bf Brane vertex operators in NSR superstring.}  
 NSR critical string theory includes a 
specific class of NS physical states (BRST invariant and BRST
nontrivial
vertex operators)
which are not associated with any perturbative particle
emission but appear to play an important role in the non-perturbative
physics. For example in the previous works we have shown that
inserting   this new class of operators to NSR
 string theory is 
somewhat equivalent to introducing branes;
In particular, some of these operators dynamically deform flat
ten-dimensional space time to $AdS_5\times{S^5}$ background
\refs{\myself}

Generally, these states are described by vertex operators
that exist at nonzero ghost pictures only, not admitting ghost
number zero representation. This crucially distinguishes them from
usual perturbative open or closed string states, such as
photon, graviton or dilaton, which in principle are allowed to
exist at any arbitrary ghost picture.
This new class of states appears in both closed and open superstring
theories and includes both massless and tachyonic states.
In open string theory the massless ones are represented by two-form
and five-form vertex operators;
they are given by:

\eqn\grav{\eqalign{V_{m_1...m_5}^{(-3)}(z,k)=
e^{-3\phi}\psi_{m_1}...\psi_{m_5}e^{ikX} \cr
V_{m_1...m_5}^{(+1)}(k)=e^{\phi}\psi_{m_1}...\psi_{m_5}e^{ikX}(z)
+ghosts\cr
V_{m_1m_2}=e^{-2\phi}\psi_{m_1}\psi_{m_2}e^{ikX}}}
These dimension 1 primary fields must also be integrated over the 
worldsheet boundary
The five-form picture $-3$
operator $V^{(-3)}_{m_1...m_5}$ is  
BRST-invariant at any momentum, as is easy to see by simple and
straightforward computations involving the BRST charge given
by
$$Q_{brst}=\oint{{dz}\over{2i\pi}}
c(T_{matter}+{1\over2}T_{ghost})+\gamma(G_{matter}+{1\over2}G_{ghost})$$
which may also be written as
$$Q_{brst}=\oint{{dz}\over{2i\pi}}{\lbrace}cT-{1\over2}e^{\phi-\chi}
\psi_m\partial{X^m}-{1\over4}be^{2\phi-2\chi}+b:c\partial{c}\rbrace$$
with T being the full matter $+$ ghost stress-energy tensor
of NSR string theory.
 However   $V^{(-3)}_{m_1...m_5}$  is BRST non-trivial only if its momentum k
is polarized along the 5 out of 10 directions orthogonal
to $m_1$,...$m_5$ (for any given polarization choice of $m_i$).
Indeed, it is easy to see that the only BRST triviality
threat for the 5-form operator may appear from the expression
$\lbrace{Q_{brst},S_{m_1...m_5}}\rbrace$
where 
$$S_{m_1...m_5}=e^{\chi-4\phi}\partial\chi(\psi\partial{X})
\psi_{m_1}...\psi_{m_5}e^{ikX}$$
 - indeed, when $S_{m_1...m_5}$ is primary field of
dimension 1 (note that this is the case only if there are no 
X's with coincident indices in $(\psi_s\partial{X^s})$ and
$e^{ikX}$, i.e. the index $s$ and the momentum polarization
are chosen so that
the $(\psi_s\partial{X^s})$ is directed orthogonally
with respect to both $e^{ikX}$ and fermions
$\psi_{m_1},...\psi_{m_5}$)
we have
$$\oint{{dz}\over{2i\pi}}{cT(z)S_{m_1...m_5}(w)}\oint{{dz}\over{2i\pi}}
{1\over{z-w}}(c\partial{S_{m_1...m_5}}+\partial{c}S_{m_1...m_5})
=\partial(S_{m_1...m_5})$$
i.e. the full derivative  which vanishes  after integrating
the vertex operator
over the worldsheet boundary.
Next, obviously, there is no pole in the O.P.E. between
$\oint{b:c\partial{c}}$ of $Q_{brst}$ and $S_{m_1...m_5}$
(since S contains no fermionic ghosts) and
also there is no pole in the O.P.E. between
$\oint{e^{2\phi-2\chi}b}$ of $Q_{brst}$ and $S_{m_1...m_5}$ since
$$e^{2\phi-2\chi}b(z)e^{\chi-4\phi}\partial
\chi(\psi\partial{X})
\psi_{m_1}...\psi_{m_5}e^{ikX}(w)\sim $$ 
$$ (z-w)
be^{-2\phi-\chi}(\psi\partial{X})\psi_{m_1}...\psi_{m_5}e^{ikX}+O(z-w)^2 
$$
and therefore these terms do not contribute to the
commutator of S with the BRST charge.
But the commutator of $S$ with
the $\oint{\gamma(\psi\partial{X})}$ 
in the BRST  charge does give the
$V_5$ operator as the relevant O.P.E. has a simple pole:
$$e^{\phi-\chi}(\psi\partial{X})(z)e^{\chi-4\phi}\partial\chi
(\psi\partial{X})\psi_{m_1}...{\psi_{m_5}}e^{ikX}(w)\sim
{{1}\over{z-w}}e^{-3\phi}\psi_{m_1}...\psi_{m_5}$$
and therefore we have the BRST triviality
$$\lbrace{Q_{brst},S_{m_1...m_5}}\rbrace{\sim}e^{-3\phi}
\psi_{m_1}...\psi_{m_5}$$
 However, it is clear that if the momentum
$k$ is orthogonal to $m_1,...m_5$ directions 
the $S_{m_1...m_5}$ is not a primary field:
the $(\psi\partial{X})$ part of it always has internal O.P.E.
singularities with either $\psi_{m_1}...\psi_{m_5}$
or $e^{ikX}$.
 As a result, whenever the momentum k is
orthogonal to $m_1,...m_5$ directions,
the O.P.E. of the stress-energy tensor
with $S_{m_1...m_5}$ always has a cubic singularity. Therefore
$S_{m_1...m_5}$ does not commute with the $\oint{cT}$ term
of $Q_{brst}$ and 
$\lbrace{Q_{brst},S_{m_1...m_5}}\rbrace$ does not reproduce the 
5-form vertex operator $V^{(-3)}_{m_1...m_5}$
However, in case if the momentum $k$ of $V^{(-3)}_{m_1...m_5}$
is longitudinal, i.e. is polarized along 
$m_1...m_5$ directions it is easy to see that the vertex operator
becomes BRST trivial: indeed,
it can be written as a BRST commutator with the primary field:
$\lbrace{Q_{brst},C_{m_1...m_5}}\rbrace$
with
$$
C_{m_1...m_5}=e^{\chi-4\phi}\partial\chi(\psi\partial{X})^{\perp}
\psi_{m_1}...\psi_{m_5}e^{ikX}
$$
with the supercurrent part $(\psi\partial{X})^{\perp}$
now polarized orthogonally to $m_1,...m_5$, i.e.
 both to $e^{ikX}$ and other world-sheet fermions 
of the 5-form.

 So we see that BRST non-triviality condition
imposes significant constraints on the propagation of the 
5-form: namely, it is allowed to propagate in the
5-dimensional subspace transverse to its own polarization.
This also is an important and remarkable distinction of
this vertex operator
from usual vertices we encounter in perturbative string theory;
it is well known that those are able to propagate in entire
ten-dimensional space-time. 
The two-form is also BRST-invariant at any k, as is easy to
check using the above expression for $Q_{brst}$
 It is BRST-trivial at zero momentum as it can be represented as
a commutator 
$$
\lbrace{Q_{brst},e^{\chi-3\phi}\psi_{\lbrack{m_1}}\partial{X}_{m_2{\rbrack}}
\partial\chi
\rbrace}
$$
 but it becomes BRST non-trivial at non-zero momenta
and again, in complete analogy with the 5-form case,
its momentum  must be orthogonal to the $m_1,m_2$ two-dimensional
subspace, i.e. the two-form propagates in eight transverse dimensions.
Constructing the BRST-invariant version of the five-form
at picture $+1$
is a bit more tricky since the straightforward generalization
given by  $e^\phi\psi_{m_1}...\psi_{m_5}$
does not commute with two terms in the BRST current
given by $b\gamma^2$ and $\gamma{\psi_m\partial{X^m}}$.
To compensate for this non-invariance one has to add  two
counterterms, one proportional to the fermionic 
ghost number 1 field ${c}$ and another to the ghost
number $-1$ field b.

To construct these ghost counterterms one has 
to take the fourth power of picture-changing operator
$\Gamma^4\sim{:e^{4\phi}G\partial{G}\partial^2G\partial^3G:}$
with G being the full matter $+$ ghost world-sheet supercurrent
and calculate its full O.P.E. (i.e. including all the non-singular terms)
 with the picture $-3$ five-form operator.
If the $-3$-picture vertex operator is at the point 0 then
\eqn\grav{\eqalign{
V_5^{(+1)}(0)=e^{\phi}\psi_{m_1}...\psi_{m_5}-{1\over2}
lim_{z\rightarrow{0}}{\lbrace}{z^2}be^{2\phi-\chi}\psi_{m_1...m_5}\cr-{1\over2}
{z^2}ce^\chi\psi_{m_1}...\psi_{m_4}(\psi_{m_5}(\psi_n\partial{X^n})
+\partial{X_{m_5}}(\partial\phi-\partial\chi))+
O(z^3)}}
This  operator is  BRST invariant by construction since both
$\Gamma^4$  and picture $-3$ 5-form operator are BRST invariant.

The BRST commutator with counterterms must be computed at a point $z$
and then the limit $z\rightarrow{0}$ is to be taken.
Fortunately, due to the condition of fermionic ghost number conservation
this unpleasant non-local ghost part is unimportant in
computations of correlation functions and can be dropped at least
in cases when not more than one picture $+1$-operator is involved.
 For our purposes in this paper this shall be sufficient
and we will shall drop the non-local part elsewhere.
The origin of these exotic 5-form and   2-form operators 
is in fact closely related to Ramond-Ramond states at non-canonical pictures.

 Consider   the  Ramond-Ramond vertex operators at zero momentum 
in $(-1/2,-1/2)$ and $(-3/2,-3/2)$-pictures
 on a disc:
\eqn\grav{\eqalign
{V^{(-1/2,-1/2)}(k,z,\bar{z})
=e^{-{1\over2}\phi}\Sigma^\alpha(z)
e^{-{1\over2}{\bar\phi}}\bar\Sigma^\beta(\bar{z})e^{ikX}(z,\bar{z})
{\Gamma^{m_1...m_q}_{\alpha\beta}}F_{m_1...m_q}(k)\cr
V^{(-3/2,-3/2)}(k,z,\bar{z}) =
e^{-{3\over2}\phi}\Sigma^\alpha(z)
e^{-{3\over2}{\bar\phi}} \times
\bar\Sigma^\beta(\bar{z})e^{ikX}(z,\bar{z}){\Gamma^{m_1...m_q}_{\alpha\beta}}
F_{m_1...m_q}(k)}}

where $\phi(z)$ and $\chi(z)$ are free fields that appear in the
bosonization of the NSR superconformal ghosts $\beta$,$\gamma$;
$\Sigma^\alpha$ is  spin operator for  NSR matter fields.

The crucial (and often neglected) point is that
if a Ramond-Ramond vertex is placed on a disc 
and the boundary is present, the holomorphic and anti-holomorphic
matter and ghost spin operators are no longer independent but
they are related as:
\eqn\grav{\eqalign{\bar\phi(\bar{z})=\phi(\bar{z}),
\bar\chi(\bar{z})=\chi(\bar{z})\cr
\bar\Sigma^\alpha(\bar{z})={M^{(p)\alpha}_\beta}\Sigma^\beta(\bar{z})\cr
M_{\alpha\beta}\equiv({\Gamma^0...\Gamma^p})_{\alpha\beta}}}
The expression for the matrix $M^{(p)}_{\alpha\beta}$
implies that the Dirichlet boundary conditions are
imposed on p out of 10 $X^m$'s while the Neumann conditions
are imposed on the rest.
As long as  the vertices (1) are far from the edge of a D-brane
(that is, $z\neq{\bar{z}}$) one may neglect the interaction
 between holomorphic and anti-holomorphic spin operators;
however, as one approaches the boundary of the disc where $z=\bar{z}$
the internal normal ordering must be performed inside the 
Ramond-Ramond vertex operators in order to remove the singularities
that arise in the O.P.E. between the spin operators located
at $z$ and $\bar{z}$. 
Adopting the notation
${\not{F}}^{(q)}(k)\equiv{\Gamma^{m_1}...\Gamma^{m_q}}F_{m_1...m_q}(k)$
we find that the result of the normal ordering is given by:
\eqn\grav{\eqalign{lim_{{z,\bar{z}}\rightarrow{s}}
:e^{-{3\over2}\phi}\Sigma_\alpha:
(z){{\not{F}}^{(q)}_{\alpha\beta}}:e^{-{3\over2}\bar\phi}\bar\Sigma_\beta
(\bar{z}):\cr\sim{1\over{z-\bar{z}}}Tr({{\not{F}}^{(q)}}M^{(p)}
\Gamma^{m_1...m_5})e^{-3\phi}\psi_{m_1}...\psi_{m_5}(s)+...\cr
{lim_{{z,\bar{z}}\rightarrow{s}}}
:e^{-{3\over2}\phi}\Sigma_\alpha:
(z){{\not{F}}^{(q)}_{\alpha\beta}}:e^{-{1\over2}\bar\phi}\bar\Sigma_\beta
(\bar{z}):\cr\sim{1\over{z-\bar{z}}}Tr({{\not{F}}^{(q)}}M^{(p)}
\Gamma^{m_1m_2})e^{-2\phi}\psi_{m_1}\psi_{m_2}(s)+...}}
where we have dropped the less singular terms in the O.P.E.
as well as full derivatives.
We see that due to the internal normal ordering at the
boundary of the disc the
Ramond-Ramond vertex operators degenerate into massless
 $open-string$ vertices - the two-form 
$Z_{mn}=e^{-2\phi}\psi_m\psi_n$ and the five-form
$Z_{m_1...m_5}=e^{-3\phi}\psi_{m_1}...\psi_{m_5}$.
 Giving a proper physical interpretation to these
$new$ massless states in the spectrum of a superstring obviously
is a challenging puzzle. In our previous works
\refs{\dimacohomology} we have also
 shown that two-form and five-form
vertices (3)  appear as central terms in the 
space-time superalgebra for NSR superstring theory when the
supercharges are taken in non-canonical pictures.
The proof is quite analogous the above
    derivation of the
brane-like states from the Ramond-Ramond insertions on a disc.
Since p-form central terms in a SUSY algebra are always related to 
the presence of p-branes, this leads us to conjecture
that the 2-form and 5-form operators are related to non-perturbative
dynamics of branes in string theory; so to speak they may be thought
of as vertex operators creating extended soliton-like objects
(unlike usual string vertex operators that correspond
to emission of point-like particles)

The brane-like states (1) also 
have their analogues in the closed string-sector.
To construct the closed string counterpart of the 5-form state (1)
let us split the ten-dimensional space-time 
index $m$ in the 4+6 way:
$X_m=(X_a, X_t)$
where  $m=0,...9 ; a=0,...3; t=4,...9$
and similarly for the worldsheet fermions.
Then the relevant  closed string vertex operator is
given by:
\eqn\grav{\eqalign{V_5(z,\bar{z},k^{||})=\lambda(k^{||})
\epsilon_{a_1...a_4}e^{-3\phi-{\bar\phi}}
\psi_{a_1}...\psi_{a_4}\psi_t{\bar\psi^t}e^{ik^{||}X}(z,\bar{z})\cr
V_5^{*}(z,\bar{z},k^{||})=\lambda(k^{||})
e^{-3\phi-\bar\phi}\psi^{{\lbrack}t_1}...\psi^{t_5}
\bar\psi^{t^6\rbrack}e^{ik^{||}X}\epsilon_{t_1...t_6}}}
with $k^{||}X\equiv{k_aX^a}$
which we will also refer to as $V_5$-operator in the rest of the paper.
The $V^{*}$ operator describes excitations of a D5-brane in
4 directions transverse to its  worldvolume,
while the $V_5$-operator effectively corresponds to a D3-brane
(obtained by wrapping the 5-brane along the 2-cycle) and
its excitations are confined to its 4-dimensional
worldvolume. An important remark should be made here to
avoid possible confusions: 
note that for the sake of compactness
in our  formulae we chose the worldvolume of a D5-brane
to span the 6 directions $t_1,...t_6$ which on the other
hand correspond to
transverse coordinates of a D3-brane in our definition of
$V_5$; but of course it should be understood that the
physical implications of the four-dimensional momentum
$k^{||}$
in $e^{ik^{||}X}$ are significantly different
in cases of $V_5$ and $V_5^{*}$ : in the first case it
corresponds to longitudinal oscillations in the D3-brane
worldvolume while in the second case ($V_5^{*}$) $k^{||}$
accounts for transverse oscillations of the $D5$-brane,

The properties of the sigma-model with the $V_5$-operator have been studied
in ~\refs{\myself} where the relevance of this vertex operator
to non-perturbative $D3$-brane  dynamics has been shown
(both $V_5$ and $V_5^{*}$ operators can be used in the sigma-model
in an equivalent way)
It is important that BRST invariance condition for the $V_5$
and $V_5^{*}$-operators
(3) requires that their propagation is confined to four dimensions.
Namely, to insure the BRST invariance,
the momentum $k^{||}$ must be polarized along the  $a_1,...a_4$
directions.
Indeed, it is easy to see that the only way to avoid a cubic
singularity in the O.P.E. between the antiholomorphic BRST
current and $V_5$ (which arises from the $\bar{c}{\bar\partial}X_m
\bar\partial{X^m}$ term in $j_{brst}$ and destroys the BRST invariance
of $V_5$)
one has to take the $X's$ in the exponent of $V_5$
orthogonal to the  ${{X^t}}$ in the antiholomorphic part,
i.e. the momentum should be polarized along the longitudinal
$a=0,...3$ four dimensional subspace; for the $V_5^{*}$
operator everything goes totally analogously.
Again, as in the case of open string theory one is able to show BRST 
non-triviality of $V_5^{*}$ and $V_5$.
Indeed, the only possible
 possible BRST triviality threat is once again coming from the 
commutator of $\int{\gamma(\psi\partial{X})}$ in
holomorphic part of $Q_{brst}$
with $C=e^{\chi-4\phi-\bar\phi}\psi^{{\lbrack}t_1}...\psi^{t_5}
\bar\psi^{t_6\rbrack}(\psi\partial{X})\partial\chi{e^{ik^{||}X}}$
but again C is not a primary field, having a cubic
O.P.E. singularity with stress-energy tensor
and therefore its commutator with $Q_{brst}$ does not 
reproduce $V_5^{*}$ - and similarly for $V_5$.
Moreover, the condition of the world
sheet  conformal invariance
(preserving the conformally invariant form of the O.P.E. between
two stress-energy tensors corresponding to the action (2) or
the vanishing of the beta-function in the lowest order of string
perturbation theory) requires that the space-time scalar field
$\lambda(k^{||})$, corresponding to the $V_5$-operator, should behave as
\eqn\lowen{\lambda(k^{||})\sim{{\lambda_0}\over{{{k^{||}}^4}}}}
where $\lambda_0$ is constant.
The $V_5$-operator has manifest $SO(1,3)\times{SO(6)}$
isometry and therefore  NSR sigma-model with the $V_5$ 
 operator has the same space-time Lorenz symmetry as the
 Green-Schwarz action 
of string theory on $AdS_5\times{S^5}$ with the gauge  kappa-symmetry
fixed. 
Indeed, as it has been argued in \refs{\myself},
 the role of the $V_5$ operator
is that it transforms the flat ten-dimensional space-time vacuum
into that of $AdS_5\times{S^5}$, thus connecting two maximally supersymmetric
backgrounds in ten dimensions. This is because  adding the $V_5$-term to
the sigma-model action (2) is in fact equivalent to introducing  $D3$-branes
in the theory. As a result, one may explore string
theory in the AdS background (and  consequently,
 the large N limit of gauge theory) by means of the brane-like
sigma-model (2) which technically lives in $flat$ ten-dimensional 
space-time. 
Using the AdS/CFT correspondence
\refs{\malda, \ampf,  \wit }, i.e. the
correspondence  between local gauge invariant
operators in the large N Yang-Mills theory and 
massless vertex operators in string theory one may obtain
the large N correlators in gauge theory by computing  
scattering amplitudes of the BRST invariant vertices in the 
sigma-model (2).  For example, as the dilaton vertex operator $V_\varphi$
corresponds to the $Tr{F^2}$ field in gauge theory, the generating
functional for various correlation functions of the $Tr{F^2}$ 
operators is given by:
\eqn\grav{\eqalign{Z(\lambda_0,\varphi)=
\int{D}\lbrack{X}\rbrack{D}\lbrack\psi\rbrack{D}\lbrack{ghosts}\rbrack
f(\Gamma,N){exp}\lbrace\int{d^2}z
\partial{X^\mu}\bar\partial{X_\mu}+\psi^\mu\bar\partial\psi_\mu
+\bar\psi^\mu\partial\bar\psi_\mu\cr+\lambda_0\epsilon_{a_1a_2a_3a_4}
\int{{d^4{k^{||}}}\over{{k^{||}}^4}}
e^{-3\phi}\psi^{a_1}\psi^{a_2}\psi^{a_3}\psi^{a_4}\psi_t\bar\partial{X^t}
e^{ik^{||}X}+\int{d^{10}p}V_\varphi(p,z,\bar{z})\varphi(p)\rbrace
\cr+c.c.+ghosts}}
where $\varphi(p)$ is ten-dimensional space-time dilaton field.
The ``measure function'' $f(\Gamma,N)\sim{(1+N^2\Gamma^4)^{-1}}
\times{c.c.}$
( $\Gamma$ is picture-changing operator and $N$ corresponds
to the gauge group parameter) needs to be introduced
to the measure of integration to insure correct ghost number balance
on the sphere
and normalization of scattering amplitudes.
The two-point dilaton correlation function, corresponding to
the  generating functional (5) is given by
\eqn\grav{\eqalign{<V_\varphi(p_1)V_\varphi(p_2)>_{\sigma-model}
={{\delta^2{Z(\lambda_0,\varphi)}}\over{\delta\varphi(p_1)\delta\varphi(p_2)}}
|_{\varphi=0}}}
To compute this correlator we have to expand the functional (5)
in $\lambda_0$. The first non-trivial contribution has the order
of $\lambda_0^2$ and it is given by
\eqn\grav{\eqalign{A_{\lambda_0^2}(p_1,p_2)\sim
\lambda_0^2\int{{d^4{k_1^{||}}}\over{{k_1^{||}}^4}}
\int{{d^4{k_2^{||}}}\over{{k_2^{||}}^4}}<V_\varphi(p_1)V_\varphi(p_2)
V_5(k_1^{||})V_5(k_2^{||})>}}
where the four-point amplitude should be computed in the usual
NSR string theory in flat space-time.
In other words, this is just the usual four-point closed string
Veneziano amplitude which has to be integrated over internal momenta
of the $V_5$-vertices, i.e. over two out of three independent momenta.
The straightforward computation of the four-point amplitude and 
the integration over the $V_5$ momenta has been performed in ~refs{\self}
and the answer is given by:
\eqn\grav{\eqalign{A_{\lambda_0^2}\sim
\lambda_0^2{(p_1^||)^4}{log}(p_1^{||})^2\int
{d^2}w{{{log(|log|w||)}+log(|log|1-w||)}\over{|1-w|^4}}}}
where $p_1^{||}$ is the longitudinal projection of the dilaton
momentum to four longitudinal directions;
$(p^||)^2=p_\alpha{p}^\alpha$. It is remarkable that the amplitude
(8) depends exclusively on four-dimensional longitudinal projection
of the dilaton momentum; up to normalization it  has the same form
as the two-point correlator $<Tr{F^2}(p_1^{||})Tr{F^2}(-p_1^{||})>$
in the $N=4$ super Yang-Mills theory in $D=4$, computed
in the  approximation of dilaton s-wave ~\refs{\ampf}.
 Fourier transforming  the amplitude (8), one recovers
the well-known expression for the two-point
amplitude in the $N=4$ $D=4$  SYM theory in the four-dimensional
coordinate space:
$Tr{F^2}(x)Tr{F^2}(y)\sim{1\over{|x-y|^8}}$.
Furthermore, the momentum structure of  amplitudes with
more $V_5$ insertions agrees with the form of the
$<Tr{F^2}Tr{F^2}>$ correlators computed at higher values
of the dilaton angular momentum; in other words,
expansion in the $\lambda_0$ parameter in the
brane-like  sigma-model (2)  accounts for
higher partial waves of the dilaton field in the 
$AdS_5\times{S^5}$ supergravity.
Proceeding similarly, one can in principle  compute higher point
correlation functions  from the generating functional
(5) to show their agreement with the known expressions
for 3 and 4-point correlators in the $N=4,D=4$ SYM theory.

To explore the mechanism of the dynamical compactification
of flat ten-dimensional space-time  on $AdS_5\times{S^5}$
due to presence of the $V_5$  vertex in the
 sigma-model  action one has to study the  modification of the dilaton's
beta-function  in the $V_5$-background.
Such an analysis has been carried out in \refs{\myself}.
 The analysis of the dilaton's beta-function
shows that the compactification 
on $AdS_5\times{S^5}$ occurs
as a result of  certain very special non-Markovian stochastic process.
 Namely, 
 the $V_5$ background in the sigma-model
 has a meaning of a ``random force'' term with the $V_5$-operator
playing the role of a non-Markovian stochastic noise, 
which correlations are determined
by the worldsheet  beta-function associated with the $V_5$ vertex.
 Indeed, The straightforward computation shows that     
the dilaton's  beta-function equation
in the presence of the $V_5$-term
has the form of the non-Markovian Langevin equation:

\eqn\lowen{
{{d\varphi(p)}\over{d(log\Lambda)}}=
-\int{d^{10}q}C_\varphi(q)\varphi({{p-q}\over2})\varphi({{p+q}\over2})
+\eta_{5}(p^{||},\Lambda)}
where
\eqn\lowen{\eta_{5}(p^{||},\Lambda)\equiv
-\lambda_0^2
(1+\lambda_0\int{d^4k_2^{||}}
\int_0^{2\pi}d\alpha\int_0^\infty{dr}rV_5(r+\Lambda,\alpha,k^{||}))}
In this equation
 the role of the stochastic noise term being played by 
the truncated worldsheet integral of the $V_5$-vertex.
The logarithm of the
worldsheet  cutoff parameter plays the role of the stochastic
time in the  Langevin equation.
The noise is non-Markovian and it is generated by
 the $V_5$ operator , as was already noted above.

The noise correlations in stochastic time are given by 
the worldsheet correlators of the $V_5$ vertices
(one has to take their worldsheet integrals at different
cutoff values and to compute  and to evaluate the cutoff dependence)
Knowing the $V_5$-noise correlators it is then straightforward to
derive the corresponding non-Markovian Fokker-Planck
equation for this stochastic process and to show that the
Fokker-Planck distribution solving this equation is given by
 the exponent of the ADM-type  $AdS_5$ gravity Hamiltonian
(computed from the $AdS_5$ gravity action at a constant
radial AdS ``time'' slice using the Verlinde's prescription
\refs{\verlinde}. Such a mechanism
 naturally relates
the radial AdS coordinate, stochastic time and the
worldsheet cutoff, pointing out an intriguing relation between
holography principle, AdS/CFT correspondence and 
non-Markovian stochastic processes.
 Therefore from space-time point of view the 
$V_5$ insertion leads to non-Markovian stochastic process
which deforms flat ten-dimensional space-time
geometry  the one of  to $AdS_5\times{S^5}$.
At the same time, from the worldsheet point of view
the situation looks as follows. In the beginning we have
a critical NSR string theory in flat  
space-time with a simple 2d conformal field theory on a worldsheet.
This $CFT$ is perturbed by the $V_5$ vertex operator and as a result
the worldsheet theory flows to some new fixed point, i.e.
new CFT.
It is this new CFT which , in agreement with the arguments above,
should constitute the worldsheet theory of NSR strings on 
$AdS_5\times{S^5}$.
 Also
one can consider the NSR theory perturbed by
 open string $V_2$-operator (two-form) which effectively corresponds to a 
D-string propagating in 8 dimensions (transverse
to its worldsheet). Its $SO(1,1)\times{SO(8)}$ covariant
form should be given by
$$\lambda(k^{\perp})\epsilon_{ab}e^{-2\phi}\psi_a\psi_b{e^{ik^{\perp}X(z)}}$$.

 An important problem to consider is to check
(by analyzing appropriate correlation functions in this sigma-model)
 that the $V_2$ operator effectively curves the background
to give the $AdS_3$ compactification (just like the $V_5$-perturbation
gives us $AdS_5\times{S^5}$). If the answer is positive  that
would mean that we can generate all the essential $AdS$ backgrounds
in ten-dimensional superstring theory
by merely perturbing the flat space-time  theory by
 the $V_5$  and $V_2$ pair.To complete a brane zoology in terms of
brane-like vertex operators one also  needs to construct
 a closed string  version of the brane-like two-form.
  The construction is completely analogous
to the $V_5$ case and the BRST invariant closed-string
partner of $V_2$ is
given by:
\eqn\grav{\eqalign{
{V_2^{*}
\sim\rho(k^{\perp})\epsilon^{\alpha\beta\gamma}e^{-2\phi-\bar\phi}
\psi_\alpha\psi_\beta\bar\psi_\gamma{e^{ik^{\perp}X}}}\cr
V_2\sim{\rho_\alpha(k^{||})} e^{-2\phi-\bar\phi}\psi_\alpha
\psi_t\bar\psi^t e^{ik^{||}X}  }}
(9 indices t are orthogonal to $\alpha$ in $V_2$,
$k^{||}\equiv{k_\alpha}$).The $V_2$ operator describes the 
$D0$-brane whose momentum is directed along a given $\alpha$ direction
($\alpha$ also labels the 1-dimensional D0-brane
worldline along which the momentum is polarized)
while the dual
$V_2^{*}$ vertex should account for the membrane
(with $\epsilon^{\alpha\beta\gamma}$ spanning its
three-dimensional worldvolume)

To summarize, we have the following classification:

D0-brane is described by the closed string $V_2$-vertex;

D1-brane by the open-string two-form ;

D2-brane by the closed string $V_2^{*}$-vertex;

D3-brane by the closed string $V_5$-vertex;

D4-brane by the open-string five-form;

D5-brane by the closed string $V_5^{*}$ vertex.

Let us ask now a natural question - we know that even branes, i.e.
$D0, D2, D4$ branes exist in type A theory (either type IIA or $0$A) and 
 odd branes, i.e. $D1, D3, D5$ exist in type B theory. It is well known 
that difference between types A and B is  due to fermion numbers in
left and right Ramon sectors \refs{\jp} -
 namely type A has $(R+, R-)$ or $(R-, R+)$ or both of them (for type
$0$A) and type B has either $(R+,R+)$ or $(R_,R_)$ or both of
them.  But by direct inspection of $V_2$ and $V_2^{*}$ one can see
that the difference between  numbers of left and right fermions is odd
(namely one)  and  open-string five-form 
vertex operator also have odd (namely
five) fermions - so they all must  create branes in type A theory -
and indeed in our table they correspond to even branes. By direct
inspection of  $V_5$ and $V_5^{*}$  one can see
that the difference between  numbers of left and right fermions is
even (namely four)  and  open-string two-form vertex operator 
also have even (namely  two) fermions - so they all must  create
branes in type B theory -
and indeed in our table they correspond to odd branes.

 For example to get $D1-D5$ system we have to deform our sigma model
by closed string $V_5^{*}$ and open string two-form $V_{m_1,m_2}$
 vertex operators in which open string operator is polarized in
transverse eight directions and  $V_5^{*}$ is polarized in a
four-dimensional subspace of this eight-dimensional space. Thus
 addition of $V_5^{*}$ vertex   will deform symmetry group 
$SO(1,1)\times{SO(8)}$  we have discussed earlier down to 
$SO(1,1)\times{SO(4)}\times{U(1)^{4}}$ which is precisely the symmetry 
group of $D1-D5$ system. Of course in the near-horizon limit it
corresponds to the $AdS^3 \times S^3 \times T^4$ metric \refs{\malda}.

This completes our discussion  of
non-perturbative brane-like vertex
operators.
 In the next section  we shall attempt to analyze some
properties of the new worldsheet CFT created by brane-like vertex
operators.

\centerline{\bf NSR AdS Strings and Logarithmic Operators}

Sometime ago \refs{\KM} it was suggested that world sheet dynamics describing
backgrounds with collective coordinates must be described by
Logarithmic Conformal Field Theory (LCFT) 
\refs{\Gur}. The arguments were based on  hidden symmetries in LCFT
and an existence of a logarithmic zero norm state  \refs{\CKT}.  
In  \refs{\KMW} a logarithmic pair
  describing  D-brane recoil  was constructed explicitly. The
logarithmic recoil operator (so called D operator)
 looks very similar to operator $V_5$ integrated over momenta.
 So it seems reasonable to suggest that we may deal with LCFT here
too.  There is another reason to suspect that   worldsheet
dynamics is given by LCFT in this theory. As we saw the brane vertex
operators has a very unusual feature - they mix matter and ghost
fields. It was suggested recently \refs{\susy}
            that  in theories with
matter-ghost mixing  one has LCFT on a world sheet. 

To  prove that we have LCFT let us consider
 the following pair of operators:
\eqn\grav{\eqalign{L_5=\int{{d^4p}\over{p^4}}e^{-3\phi-\bar\phi}
\psi_0...\psi_3\psi_t\bar\psi^t{e^{ipX}}(z,\bar{z});\cr
N_5=\int{{d^4k}\over{k^2}}e^{\phi-\bar\phi}
\psi_0...\psi_3\psi_t\bar\psi^t{e^{ikX}}(w,\bar{w})}}
The operator product expansion of $L_5$ with itself is given by: 

\eqn\grav{\eqalign{L_5(z,\bar{z})L_5(w,\bar{w})\sim\cr
\int{{d^4pd^4q}\over{p^4q^4}}(pq){1\over{|z-w|^2}}
e^{-2(pq)log|z-w|}V_\varphi^{(-2,-2)}(p+q)}}
where $V_\varphi$ is dilaton vertex operator in the $(-2,-2)$ picture:
\eqn\grav{\eqalign{V_{varphi}\sim{e^{-2\phi-2\varphi}}
\partial{X_m}\bar\partial{X_n}
(\eta^{mn}-k^m\bar{k^n}-\bar{k^m}{k^n})\cr
k^2=\bar{k}^2=0;(k\bar{k})=1}}

In principle this O.P.E. also contains a quartic pole proportional to
$\sim{{V_{T}}\over{|z-w|^4}}$
where $V_T=e^{-2\phi-2\bar\phi}e^{ik^{||}X}$ is BRST-invariant operator
of a clearly tachyonic nature but fortunately  it is BRST trivial
since
\eqn\lowen{e^{-2\phi}e^{ik^{||}X}\sim\lbrace{Q_{brst},e^{\chi-3\phi}\partial
{\chi}(k^{||}\psi)e^{ik^{||}X}}\rbrace}
and therefore can be dropped elsewhere. 
So the above O.P.E. 
may be written also as
\eqn\grav{\eqalign{{1\over{|z-w|^2}}{{\partial}\over{\partial{log}|z-w|}}
\int{{d^4pd^4q}\over{p^4q^4}}e^{-2(pq)log|z-w|}V_\varphi^{(-2,-2)}
(p+q)}}

Performing the change of variables in the momentum space:
$l=1/2(p+q),k=1/2(p-q)$ we write the integral as
\eqn\grav{\eqalign{L_5L_5\sim{{1}
\over{|z-w|^2}}{{\partial}\over{\partial{log}|z-w|}}
\int{d^4l}V_\varphi(l)\int{{d^4k}\over{(k-l)^4(k+l)^4}}
e^{(k^2-l^2)log|z-w|}\cr\equiv{1\over{|z-w|^2}}
{{\partial}\over{\partial\alpha}}
\int{{d^4k}\over{(k-l)^4(k+l)^4}}e^{i\alpha(k^2-l^2)}}}
Now we  denoted $i\alpha\equiv{log}|z-w|$; we shall
evaluate the momentum integral at $real$ values
of the $\alpha$ parameter, performing afterwards the
straightforward analytic continuation.
Furthermore, in our evaluation of the momentum integral over
$d^4k$ it is convenient to make the translation $(k-l)\rightarrow{k}$
which does not change the Jacobian.
So we have to evaluate the integral
\eqn\grav{
\eqalign{I(l,\alpha)=
\int{{d^4k}\over{k^4(k+2l)^4}}e^{i\alpha(k^2+2kl)}
= I_1+I_2+I_3+I_4 =
\cr
\int_{-\infty}^{\infty}dk_0\int d^3{\vec{k}}
{{e^{i(k^2+2kl)\alpha}}\over
{{\lbrack}(k_0-|k|)(k_0+|k|)(k_0+2l_0-|k+2l|)
(k_0+2l_0+|k+2l|){\rbrack}^{2}}}}}

The integral in $k_0$ has four poles and $I_1,...I_4$
are corresponding residues.
The first residue, at $|k|=k_0$ gives the three-dimensional
spatial momentum integral
\eqn\grav{\eqalign{I_1(l,\alpha) = \int{d^3{\vec{k}}}\partial_{|k|}
\lbrace{{e^{i|k|(l_0-|l|cos\theta)\alpha}\over
{4(l_0-|l|cos\theta)^2|k|^4}}}{\rbrace}\cr=
-{1\over4}\int_0^\pi{d{cos\theta}}\int_0^\infty{{d|k|}\over
{|k|^3}}e^{i|k|\alpha(l_0-|l|cos\theta)}}}
where $\theta$ is the angle between the spatial
parts of the l and k vectors; $ |l|,|k|$ are absolute values of the spatial 
parts.
We used integration by parts
and the on-shell condition $l^2=0$.
Finally, integrating over |k| and using the formula
\eqn\lowen{\int_0^\infty{{dx}\over{x^3}}e^{iax}\sim{1/2a^2log(a)-a^2/4}}
we get
\eqn\grav{
\eqalign{I_1={i\over8}\alpha^2\int_0^\pi{d}(cos\theta)
\lbrace{log}(l_0-|l|cos\theta)+log\alpha-1/2\rbrace
\cr={i\over8}\alpha^2(log(l^2)+2log\alpha-1)}}
On the other hand, the second residue at $k_0=-|k|$
gives
\eqn\lowen{I_2={i\over8}\alpha^2(log(l^2)-2log\alpha+1)}
The third and the fourth residues are evaluated likewise and in fact
$I_1+I_2=I_3+I_4$, as is  easy to check.
Summing all the four residues together and performing analytic
continuation in $\alpha$ we get
\eqn\grav{\eqalign{I\equiv{I_1+I_2+I_3+I_4}={1\over4}\alpha^2log{l^2}=
{1\over4}log^2|z-w|log{l^2}.}} 
Substituting this result into the o.p.e.
between to $L_5$'s (involving the differentiation
with respect to $log|z-w|$) we get
\eqn\grav{\eqalign{{L_5(z,\bar{z})L_5(w,\bar{w})\sim
{{log|z-w|}\over{|z-w|^2}}\int{d^4k}log(k^2)V_\varphi(w,\bar{w})}}}
with $V_\varphi$ being again the dilaton vertex operator in the
$(-2,-2)$ picture.
Next, consider the second O.P.E.
$L_5N_5$ Everything goes quite similarly:
\eqn\grav{\eqalign{L_5(z,\bar{z})N_5(w,\bar{w})\sim
{\partial\over{\partial{log}|z-w|}}\int{{d^4pd^4q}\over{p^4q^2}}
e^{-2(pq)log|z-w|}{{V_\varphi(p+q)}\over{|z-w|^2}}\cr=
{i\over{|z-w|^2}}{\partial\over{\partial{log}|z-w|}}
\int{d^4l}V_\varphi(l)\int{{d^4k}\over{k^4(k+2l)^2}}
e^{ilog|z-w|(k^2+2kl)}}}
Again, splitting the momentum integral into spatial
and $dk_0$ parts and evaluating the 4 residues
at the poles in $k_0$ we find the momentum integral to give
$I=2(I_1+I_2)$ where
\eqn\grav{\eqalign{I_1=
\int_0^\pi{d{cos}\theta}\int_{0}^\infty{{d|k|}\over{|k|^2}}
(l_0-|l|cos\theta)^{-1}e^{2i|k|(l_0-|l|cos\theta)log|z-w|}\cr=
log|z-w|(log(l^2)+loglog|z-w|-1)}}
and likewise
\eqn\grav{\eqalign{I_2=log|z-w|(log(l^2)-loglog|z-w|+1)\cr
{I=I_1+I_2}{\sim}log|z-w|log(l^2)}}
Substituting into the O.P.E. we see that the logarithm
drops out of the final answer (because of differentiating over log)
and we get
\eqn\lowen{L_5(z)N_5(w)\sim{1\over{|z-w|^2}}\int{d^4l}log(l^2)
V_\varphi(l)}
Finally, consider the O.P.E. $N_5N_5$.

Proceeding exactly as above we get
\eqn\grav{\eqalign{N_5(z,\bar{z})N_5(w,\bar{w})\sim{i\over{|z-w|^2}}\int
{d^4l}{V_\varphi(l)}{\partial\over{\partial{log}|z-w|}}
\int{{d^4k}\over{k^2(k+2l)^2}}e^{i(k^2+2kl)log|z-w|}}}
Evaluating the residues in the integral over $d^4k$ as above
we see that the entire dependence on $log|z-w|$ goes away in the
final answer and the integral is proportional to
$I=I_1+I_2+I_3+I_4\sim{log(l^2)}$.Again, substituting into the O.P.E.
we get zero after differentiating over $log|z-w|$ (since $I$ does not
depend on log) .Hence $N_5(z,\bar{z})N_5(w,\bar{w})\sim{0}$
and therefore $L_5,N_5$  constitute a  pair
of logarithmic operators.

 The last check is to see that we have indeed correct OPE in LCFT 
\eqn\grav{\eqalign{T(z)L_5(w)\sim{1\over{(z-w)^2}}({L_5-N_5})+...\cr
T(z)N_5(w)\sim{1\over{(z-w)^2}}N_5+...}}
where $T$ is the full matter $+$ ghost stress tensor.
These relations are easy to check - they follow from the definitions
of $L_5$ and $N_5$ and the operator product:
\eqn\grav{\eqalign{
T(z)e^{-3\phi-\bar\phi}\psi_0...\psi_3\psi_t\bar\partial{X^t}e^{ik^{||}X}
(w,\bar{w})\cr
\sim{1\over{(z-w)^2}}(1-{{{k^{||}}^2}\over2})
e^{-3\phi-\bar\phi}\psi_0...\psi_3\psi_t\bar\partial{X^t}e^{ik^{||}X}
(w,\bar{w})\cr+
{1\over{z-w}}\partial_w
e^{-3\phi-\bar\phi}\psi_0...\psi_3\psi_t\bar\partial{X^t}e^{ik^{||}X}
(w,\bar{w})}}
It is ${k^{||}}^2$ term which cause the mixing between $L_5$ and
$N_5$ the same way as it was in a brane recoil case \refs{\KMW}.

In critical string theory we have, of course, $<TT>=0$,
therefore T has a good behaviour as a partner in a logarithmic
pair of the worldsheet LCFT \refs{\CKT}. One can assume that
 besides logarithmic $(1,1)$ pair we have discussed there must be also 
 logarithmic $(2,0)$ and $(0,2)$ pairs. The situation here is  the
same as in the two-dimensional models describing critical disorder 
 which are described by $c=0$ LCFT \refs{\disorder} in which one has a
logarithmic pair of $(2,0)$ operators. We shall discuss this issue in
a separate publication.

\centerline{\bf Matter-ghost mixing and M-theory}
 
 Let us note that  the fact that we had 2- and 5-form  vertex operator 
in string theory  is  a puzzle. We have here objects which belong to 
M-theory. So it  is very tempting to suggest that M- theory is nothing 
but  string theory with new non-perturbative brane vertex operators
included.

 However the natural question emerge - where is the extra coordinate
of M theory. It seems to us that the natural candidate is the
bosonised superconformal ghost. Indeed it looks like  very similar
to transition from D-dimensional non-critical string theory to  the
D+1 dimensional critical theory.  Liouville field is playing the role 
of an extra dimension.

 The same is going to happen here. The moment we introduced
superconformal ghosts we have another field with positive signature.
 Actually it is the only extra field in string theory which can be
interpreted as an extra dimension.  And we see that precisely brane
vertex  operators depend on it. Moreover the analysis we performed in
the introduction about the possible polarizations of momenta in brane
operators are correctly identified them with 2-and 5- branes in
M-theory.  Let us note that we also can see the exponential $\phi$
dependence is nothing like an analog of gravitational dressing in
non-critical string theory. It is interesting   fact  
that gravitational dressing  also leads to LCFT \refs{\BK}.

 So it seems that it string theory there is actually additional field
which can play the role of extra dimension - this is bosonized
superconformal ghost. Let us note that this is the only field which
can create a dimension with positive signature, because it has
positive central charge.  The picture is very similar to what we had
for the  Liouville filed when  dynamical Liouville field
played a role of a new  dimension transforming $d$-dimensional space
into $d+1$-dimensional.  What was important of course was
gravitational dressing - the fact that vertex operators depend on
Liouville field as well as background did. Otherwise it would be
sterile degree of freedom.

 The same happens when we study non-perturbative string theory. At
perturbative level we do not see superconformal ghost. Usual closed
and open string vertex operators do not depend on it. However when the 
new non-perturbative brane vertex operators are introduced we can do
it {\bf only } by making them {\bf explicitly} depending on
superconformal ghost ! This way it becomes  as important as other
coordinates.

 Let us also note that by direct inspection  all brane vertex
operators (closed as well as open) are asymmetric with respect to
superconformal ghosts from the left and right sectors! One can show
that it is impossible to construct them without introducing  different 
left and right momenta for $\phi$ and $\bar{\phi}$ - superconformal
ghosts in left and right sectors. This means that the eleventh
dimension must be {\bf compact} - another interesting prediction of
our conjecture.

The relation between superconformal ghost and the extra dimension of
M-theory may also be given sense in the context of stochastic
quantization. Indeed, in principle the
$AdS_5\times{S^5}$ geometry can be viewed as 
the infinite stochastic time  limit of solution of an infinite order
Fokker-Planck equation (which takes into account all the 
$V_5$ noise correlations
while $AdS_5$ gravity is
a solution of the Fokker-Planck equation truncated
at quadratic order
 corresponding to dilaton s-wave approximation).
 Therefore the Fokker-Planck distribution
solving
this equation is effectively eleven-dimensional away from equilibrium
point.
The role of additional dimension is played by stochastic time
but it is known that in  stochastic quantization
of gauge theories (with gauge fixing no longer necessary)
 stochastic time effectively replaces the ghost degrees of freedom
~\refs{\huffel}
At the same time, the appearance
random force term in the RG equation for the dilaton field
in the brane-like sigma-model (8)
 is closely 
to the superconformal ghost structure the $V_5$-operator,
to emphasizinghe connection between the eleventh dimension (stochastic
time) and superconformal ghost degree of freedom.

Finally, we would also like to make another comment about the
$V_5$-operator curving the background to $AdS_5\times{S^5}$.
It is known that, unlike a string theory in flat space-time,
the  $AdS_5\times{S^5}$
string satisfies the loop equation (at least in WKB approximation)
 and possesses
a zigzag symmetry which is necessary to insure confining properties of the
string. Of course the confining properties of the $AdS_5\times{S^5}$
string totally fit the context of AdS/CFT.
In this respect, the 
$AdS_5\times{S^5}$ compactification implemented by the $V_5$ operator
may be seen as a restoration of this special worldsheet zigzag symmetry
lacked by usual NSR string in flat space-time.
This restoration may be understood as follows.
It is known that necessary and sufficient condition for the
zigzag symmetry is closeness of operator algebra of  massless
 open string operators (i.e.gluons).Of course in a usual NSR model
this operator algebra is not closed as, for example, the full
O.P.E. between photons contains infinite tower of massive vertices.
However, introducing the 
$V_5$-operators should cure the open string
operator algebra in a   sense that it would enable us to 
remove these
undesired massive vertices. Namely, the operator algebra
of two
$V_5$-vertices would contain the same massive vertices but with 
opposite signs, so introducing the $V_5$-background would absorb
the massive tower of vertices in the O.P.E. algebra of photons
and therefore restore the zig-zag symmetry so that we get
the confining (i.e. $AdS_5\times{S^5}$ string theory).
In the future work we hope to examine this hypothesis of a
zigzag symmetry restoration in more details.

\centerline{\bf Discussion and  Conclusion}
 Before making a conclusion let us  make here several 
 interesting observations  about 
superconformal ghosts. The superconformal ghost $\beta -\gamma$ 
have central charge
\eqn\cbetagamma{C_{\beta\gamma} = 11}
 which is intriguing relation with the dimension of M theory.
However this is not enough - when one bosonised this system
\refs{\jp} one actually have two conformal field theories - one
describing scalar field with positive norm and central charge
\eqn\cbetagamma{C_{\phi} = 13}
and another is 
\eqn\cbetagamma{C = -2}
system of symplectic fermions.
 $C=-2$ system is an LCFT \refs{\Gur} and plays very important role in 
recently discussed critical disorded systems (see for example
 recent paper \refs{\tsvelik} and references therein). Using the fact
that  critical strings and disordered systems  are both systems with
total central charge $C_{total} = 0$ and may  have  some similarity
 \refs{\susy}  it is very interesting that  this sector is naturally 
incorporated in superstring theory.  However it did not reveal itself
so far even at level of brane vertex operators. Is it possible that at 
some other level (off shell strings ?) the $C=-2$ ghosts will be
important ? And if yes - can we also take into account $bc$ ghosts
with central charge $-26$ ?
 One can  ask an interesting question - is it possible to get  extra
dimensions from  $-26$ and $-2$ ghosts ? Obviously it will give us not 
 11-dimensional space but 12 or even 13-dimensional space, moreover
 they will have {\bf two} times and symmetry group will be $SO(10,2)$
or $SO(11,2)$ ! But  this is precisely what have been discussed recently
in relation with  F and S theories \refs{\bars} !

   It is tempting to suggest that the full non-perturbative
formulation of String theory (M,F,S, etc) is nothing but string theory
 with ALL ghosts field playing dynamical role - and with a FULL
matter-ghost mixing.

In conclusion we want to  outline several important issues which have
to be studied. First of all we have to understand how to calculate
tensions of $D$ branes in our picture and how to reproduce  all known
results about D-brane interactions. It is necessary to study in all
details the structure of BRST cohomologies in LCFT and prove unitarity 
of these theories - presence of  logarithmic zero norm states will be
very important. Relation between superconformal ghosts and Yang-Mills
ghosts which emerges   may explain amusing fact that one loop beta
functions in YM theory have coefficients proportional to conformal
anomalies in strings theory. How to get heterotic string - does the
fact that bosonised superconformal ghost has central charge  which 
is 1/2 of critical dimension of bosonic string is related that
heterotic string is half bosonic ?  Is it plausible  that  
matter-ghost mixing  may be important to understand the nature of 
famous $1<C<25$ barrier in non-critical strings ? 
 Is it possible to imagine that due 
 to stochastic description of extra dimension we shall have quantum
mixing and there will be processes in which pure state will evolve
into  mixed one ? Etc, etc, etc....

Some of these questions sounds
very strange but they definitely worth further analysis and we hope to 
return to them in future publications.

\centerline{\bf Acknowledgments}
 One of us (D.P.) would like to express his deep gratitude to
Theory group at the University of Helsinki and especially
M. Chaichian for their kind hospitality and
 Theory group at Oxford for
hospitality during a visit in December.
 The work of IIK  is
supported in part by the PPARC rolling grant PPA/G/O/1998/00567, by
the EC TMR grants  HRRN-CT-2000-00148 and  HPRN-CT-2000-00152.
D.P.   gratefully acknowledges support of 
High Energy Accelerator Research Organization
(KEK) in Tsukuba, Japan.

\listrefs
\end

\listrefs
  
\end